\documentclass{aastex63}
\shorttitle{An asymmetric dust ring around ZZ Tau IRS}
\shortauthors{Hashimoto et al.}

\graphicspath{{./}{figures/}}

\newcommand{\jh}[1]{{\color{black}#1}}

\begin{document}

\title{An asymmetric dust ring around a very low mass star ZZ Tau IRS}

\correspondingauthor{Jun Hashimoto}
\email{jun.hashimto@nao.ac.jp}

\author[0000-0002-3053-3575]{Jun Hashimoto}
\affil{Astrobiology Center, National Institutes of Natural Sciences, 2-21-1 Osawa, Mitaka, Tokyo 181-8588, Japan}
\affil{Subaru Telescope, National Astronomical Observatory of Japan, Mitaka, Tokyo 181-8588, Japan}
\affil{Department of Astronomy, School of Science, Graduate University for Advanced Studies (SOKENDAI), Mitaka, Tokyo 181-8588, Japan}

\author[0000-0001-9290-7846]{Ruobing Dong}
\affil{Department of Physics \& Astronomy, University of Victoria, Victoria, BC, V8P 1A1, Canada}

\author{Takayuki Muto}
\affil{Division of Liberal Arts, Kogakuin University, 1-24-2, Nishi-Shinjuku, Shinjuku-ku, Tokyo 163-8677, Japan}

\begin{abstract}

We present Atacama Large Millimeter/submillimeter Array (ALMA) gas and dust observations at band 7 (339~GHz: 0.89~mm) of the protoplanetary disk around a very low mass star ZZ~Tau~IRS with a spatial resolution of 0\farcs25. The $^{12}$CO~$J=3\rightarrow2$ position--velocity diagram suggests a dynamical mass of ZZ~Tau~IRS of $\sim$0.1--0.3~$M_{\sun}$. The disk has a total flux density of 273.9 mJy, corresponding to an estimated mass of 24--50~$M_\oplus$ in dust. The dust emission map shows a ring at $r=$ 58~au and an azimuthal asymmetry at $r=$ \jh{45}~au with a position angle of 135\degr. The properties of the asymmetry, including radial width, aspect ratio, contrast, and contribution to the total flux, were found to be similar to the asymmetries around intermediate mass stars ($\sim$2~$M_{\sun}$) such as MWC~758 and IRS~48. This implies that the asymmetry in the ZZ~Tau~IRS disk shares a similar origin with others, despite the star being $\sim$10 times less massive. Our observations also suggest that the inner and outer parts of the disk may be misaligned. Overall, the ZZ~Tau~IRS disk shows evidence of giant planet formation at $\sim$10 au scale at a few Myr. If confirmed, it will challenge existing core accretion models, in which such planets have been predicted to be extremely hard to form around very low mass stars.

\end{abstract}

\keywords{protoplanetary disks --- planet--disk interactions ---  planets and satellites: formation --- stars: individual (ZZ Tau IRS)}

\section{Introduction} \label{sec:intro}

Core accretion planet formation theories predict that only terrestrial and icy planets can form around very low mass (VLM) stars ($\lesssim$0.1--0.2~$M_{\sun}$), and the formation of gas giant planets is essentially prohibited \citep[e.g.,][]{liu2019a,liu2020a,matsumoto2020a,miguel2020a}. The reasons are multifold. First, a low stellar mass results in longer dynamical timescale and longer growth timescale of dust, delaying the formation of planetary cores. Secondly, the mass of dust in protoplanetary disks, the building blocks of planets, is correlated with stellar mass \citep[e.g.,][]{ansdell2017a}. A low dust mass in disks around VLM stars leads to low mass of protoplanets, suppressing pebble accretion \citep{ormel2018a} and preventing cores from ever reaching the critical mass to trigger runaway gas accretion. Last but not least, observed stellar accretion rates suggest that disks around VLM stars evolve faster than those around higher mass stars, further daunting the task of planet formation \citep{manara2012,liu2020a}. In addition, water snowline has been proposed to be the preferred site of planetesimal and planet formation due to its enhanced local dust-to-gas ratio \citep{ormel2017a}. Around VLM stars, snowline is close to the star at $r\sim$0.1~au. Therefore compact systems with rocky planets like the TRAPPIST-1 \citep{gillon2016a} are expected, while planet formation at $r\gtrsim$1~au is not \citep{miguel2020a}.

Indeed, statistical studies show that at small orbital separations rocky and icy planets are common around M~dwarfs \citep[e.g.,][]{dressing2015a}, while the occurrence rate of gas giant planets dives from 14\% around 2~M$_{\sun}$ stars to 3\% around 0.5~M$_{\sun}$ stars \citep{johnson2010}.
Around VLM stars, only a handful of gas giant planets have ever been found \citep[e.g., GJ~3512;][]{morales2019a}, and their formation mechanism is a mystery. Since their host stars are mature mean sequence stars that have finished planet formation long before, it is unclear where and when do those planets form. Protoplanetary disks around VLM stars serve as excellent laboratories for investigating planet formation in these extreme environments.

The Atacama Large Millimeter/submillimeter Array (ALMA) has revealed a variety of sub-structures in protoplanetary disks, such as rings, gaps, cavities, spirals, and crescent structures. While rings/gaps are the most common \citep[e.g.,][]{andr2018, feng2018a, cieza2020a}, roughly 10 disks show azimuthal asymmetries \citep[e.g.,][]{fran2020a,vandermarel20a,tsukagoshi2019a,perez2018a,dong2018mwc758,cieza+17}. Disks around VLM stars also harbor rings and gaps \citep[e.g.,][]{Kurtovic2020a}; however, azimuthal asymmetries have not yet been reported. Asymmetric dust crescents may be local dust traps at gas pressure maxima \citep[e.g.,][]{raettig2015,ragusa2017}, thus ideal locations of planet formation \citep[e.g.,][]{chang2010a}. Three possible origins of asymmetries have been discussed by \citet{vandermarel20a}: (a) long-lived anticyclonic vortices at gap edges, possibly opened by companions \citep[e.g.,][]{raettig2015}, (b) gas horseshoes at the edge of eccentric cavities curved by massive companions \citep[e.g.,][]{ragusa2017}, and (c) part of spiral arms \citep{vandermarel+16a} triggered by companions. The main difference between the former two is in the mass of the companions: a vortex can be produced at the edges of gaps opened by planets as low mass as super-Earths \citep{dong18doublegap}, whereas a horseshoe needs to be triggered by a much more massive companion, i.e., a brown dwarf. Overall, all three mechanisms suggest the presence of companions (planets) inside the cavity or gap.

In this paper, we report a ring and an asymmetric structure around the VLM star ZZ~Tau~IRS (spectral type: M4.5 to M5, $T_{\rm eff}$: 3015 to 3125~K, $L_{\rm star}$: 0.02 to 0.13~$L_{\sun}$, $M_{\rm *}$: 0.09 to 0.16~$M_{\odot}$, \citealp{white2004,andr2013,herczeg2014}) in the Taurus star forming region at an assumed distance of 130.7~pc\footnote{We do not use the GAIA distance of 105.7~pc because the astrometric solution is poor with a value of the renormalized unit weight error (RUWE) of 2.490 \citep[GAIA EDR3; ][]{gaia2020a}.} taken from \citet{akeson2019}. The stellar luminosity is 0.017 to 0.113~$L_{\sun}$ \citep[scaled by the updated distance]{white2004,andr2013,herczeg2014}, with the large uncertainty mainly due to factors including flux calibration and extinction (see the discussions in \citealp{herczeg2014}). With these $T_{\rm eff}$ and $L_{\rm star}$, the \citet{baraffe2015} isochrone yields a stellar mass around 0.1~$M_\odot$ at an age of 1 to 12~Myr consistently. ZZ~Tau~IRS is a single star (i.e., not identified as a spectroscopic binary; \citealp{Kounkel2019}) while its companion ZZ~Tau is 35$''$ away. Its spectral energy distribution (SED; Figure~\ref{figA:sed} in Appendix~\ref{secA:sed}) does not show any deficits at IR wavelengths $\lambda \lesssim$10~$\mu$m, and thus ZZ~Tau~IRS was not classified as having a cavity/gap structure in its disk. The large IR excess relative to the stellar flux \jh{in the SED (Figure~\ref{figA:sed})} was interpreted as an edge-on disk \citep{furlan2011}, and high-resolution optical spectroscopy of ZZ~Tau~IRS also suggested the existence of a close to edge-on disk based on its narrow emission lines from the optical jet \citep{white2004}. The total disk mass (gas~$+$~dust) was estimated to be 0.014~$M_{\odot}$ by observations at millimeter wavelengths using the Submillimeter Array \citep{andr2013,akeson2019}.

\section{ALMA archive data} \label{sec:obs}

We used the ALMA archive data at band~7 (program ID: 2016.1.01511.S, PI: J.~Patience), summarized in Table~\ref{tab:obs}. The data were calibrated by the Common Astronomy Software Applications (CASA) package \citep{mcmu07}, following the calibration scripts provided by ALMA. We performed a self-calibration of the visibilities. The phases were self-calibrated once, with fairly long solution intervals (solint = `inf') that combined all spectral windows. To facilitate analyses later (\S~\ref{sec:mcmc}), we shifted the observed dust continuum images to minimize the asymmetry relative to the phase center using the procedure \jh{described in Appendix~\ref{secA:imag}}. 

\begin{deluxetable*}{lcccc}
\tabletypesize{\footnotesize}
\tablewidth{0pt} 
\tablenum{1}
\tablecaption{ALMA observations and imaging parameters\label{tab:obs}}
\tablehead{
\colhead{Observations} & \colhead{} & \colhead{} & \colhead{} & \colhead{} 
}
%\colnumbers
\startdata
Observing date (UT)         & \multicolumn{4}{c}{2017.Jul.6}                     \\
Project code                & \multicolumn{4}{c}{2016.1.01511.S (PI: J. Patience)} \\
Time on source (min)        & \multicolumn{4}{c}{4.5}                            \\
Number of antennas          & \multicolumn{4}{c}{42}                              \\
Baseline lengths            & \multicolumn{4}{c}{16.7~m to 2.6~km}               \\
Baseband Freqs. (GHz)       & 331.9 (cont.) & 333.8 (cont.) & 343.9 (cont.) & 345.8 ($^{12}$CO~$J=3\rightarrow2$)      \\
Channel width   (MHz)       & 15.6          & 15.6          & 15.6          & 0.488      \\
Continuum band width (GHz)  & \multicolumn{4}{c}{7.5}                            \\
Bandpass calibrator         & \multicolumn{4}{c}{J0510$+$1800}                    \\
Flux calibrator             & \multicolumn{4}{c}{J0510$+$1800}                   \\
Phase calibrator            & \multicolumn{4}{c}{J0438$+$3004}                   \\
Typical PWV (mm)            & \multicolumn{4}{c}{0.60} \\
\hline \hline
\multicolumn{1}{c}{Imaging} & \multicolumn{2}{c}{Dust continuum}            & \multicolumn{2}{c}{$^{12}$CO~$J=3\rightarrow2$} \\
\hline
\jh{Robust} clean parameter      & \multicolumn{2}{c}{$-2.0$}                              & \multicolumn{2}{c}{2.0}\\
Outer $uv$-taper parameter  & \multicolumn{2}{c}{0.4~$\times$~500.0~M$\lambda$ with 48.0$\degr$} & \multicolumn{2}{c}{\nodata} \\
Beam shape                  & \multicolumn{2}{c}{254~$\times$~248~mas at PA of $-$61.7$\degr$}   & \multicolumn{2}{c}{283~$\times$~156~mas at PA of $-$50.2$\degr$}\\
r.m.s. noise (mJy/beam)     & \multicolumn{2}{c}{0.268}                                & \multicolumn{2}{c}{9.56 at 0.5~km/s bin}
\enddata
%\tablecomments{
%}
\end{deluxetable*}

The final synthesized dust continuum image of the combined data is shown in Figure~\ref{fig:dustcontinuum}. In the \verb#CLEAN# task, we set the $uv$-taper (0.4~$\times$~500.0~M$\lambda$ at PA of 48\degr) with a robust parameter of $-$2.0 to obtain a nearly circular beam, and we did not use the `multi-scale' option. The r.m.s.\ noise in the region far from the source is 268~$\mu$Jy/beam with a beam size of 254~$\times$~248~mas at a position angle (PA) of $-$61.7\degr. 

The $^{12}$CO~$J=3\rightarrow2$ line data (Table~\ref{tab:obs}) were extracted by subtracting the continuum in the $uv$ plane using the \verb#uvcontsub# task in the CASA tools. A line image cube with a channel width of 0.5~km/s was produced by the \verb#CLEAN# task. The integrated line flux map (moment~0) and the intensity-weighted velocity map (moment~1) are shown in \jh{Figure}~\ref{fig:12co32}. The channel maps at $+$1.0 to $+$13.0~km/s are shown in Figure~\ref{figA:channel} in the Appendix~\ref{secA:channel}. The r.m.s.\ noise in the moment~0 map is 61.4~mJy/beam$\cdot$km/s with a beam size of 283~$\times$~156~mas at a PA of $-$50.2$\degr$, and the r.m.s.\ noise in the moment~1 map is 9.56~mJy/beam at 0.5~km/s bin. The peak signal-to-noise ratio (SNR) is 24.1 in the channel map of $+$4.0~km/s. 

\section{Results} \label{sec:result}

\subsection{Dust continuum emission}\label{subsec:dust}

\begin{figure}[ht!]
\begin{centering}
\includegraphics[clip,width=\linewidth]{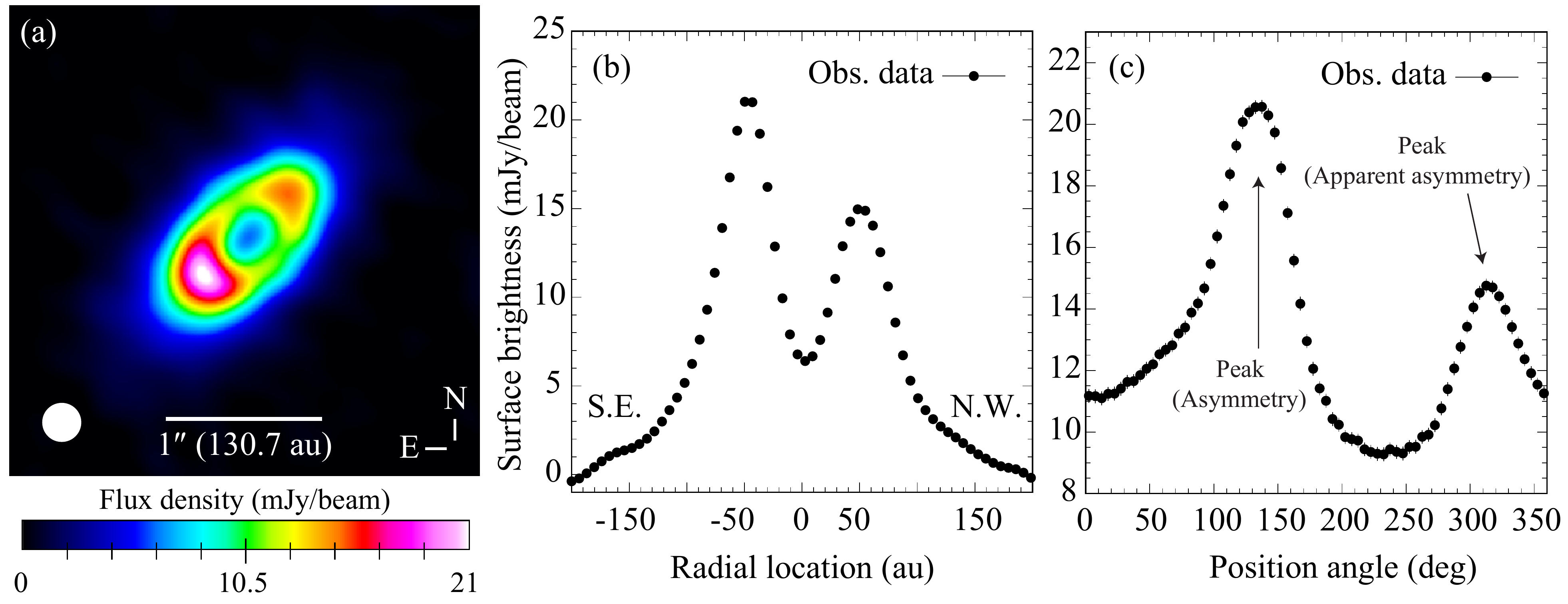}
\end{centering}
\caption{Synthesized image of the dust continuum for the ZZ~Tau~IRS ring obtained from band~7. 
(a) Entire image. The r.m.s.\ noise measured in the region far from the object is 0.268~mJy/beam with a beam size of 254~$\times$~248~mas at a PA of $-$61.7\degr. 
\jh{(b) Radial cut along the disk major axis at a PA of 135$\degr$ in panel~(a).}
\jh{(c) Azimuthal profile along the elliptical ring in panel~(a).} Two peaks are found at PAs of $\sim$130$\degr$ and \jh{$\sim$310$\degr$}. Only the peak at a PA of $\sim$130$\degr$ is a real asymmetry, whereas that at PA of \jh{$\sim$310$\degr$} is an apparent structure due to a low spatial resolution effect (see \S~\ref{sec:mcmc}).} 
\label{fig:dustcontinuum}
\end{figure}

Figure~\ref{fig:dustcontinuum} shows dust continuum images of the ZZ~Tau~IRS disk in band~7, as well as its radial and azimuthal profiles. We found two structures: a ring at $r\sim$50~au (hereafter we refer to the ZZ~Tau~IRS disk as a `ring') in Figure~\ref{fig:dustcontinuum}(b) and an azimuthal asymmetry with a contrast ratio of 1.39~$\pm$~0.05 between its peak and the opposite side on the ring in Figure~\ref{fig:dustcontinuum}(c). The peak SNR in the ring is 78.4. The dust continuum image synthesized with only the imaginary part (which efficiently detects asymmetric structures) indicates that the asymmetric structure in the ring is significant with a peak SNR of 34.3 in the central panel in Figure~\ref{figA:long} in the Appendix~\ref{secA:imag}. In the azimuthal profile (Figure~\ref{fig:dustcontinuum}c), we found two peaks at PAs of $\sim$130$\degr$ and \jh{$\sim$310$\degr$}. Such a two-peak structure along a disk major axis is commonly found in low spatial resolution images of rings ($\sim$0\farcs2--0\farcs4; e.g., \citealp{ansdell2016a}). Our visibility analyses in \S~\ref{sec:mcmc} suggest that the peak at a PA of \jh{$\sim$310$\degr$} is caused by beam effects. Hereafter we refer to the peak at a PA of $\sim$130$\degr$ as the asymmetry.

The total flux density of the dust continuum derived from visibility analyses in \S~\ref{sec:mcmc} is 273.9~$\pm$~27.4~mJy \jh{(Table~\ref{tab:mcmc})}, assuming a 10~\% uncertainty in the absolute flux calibration, consistent with previous single-disk observations (300~$\pm$~40~mJy, \citealp{difrancesco2008}). The peak brightness temperature in the ring calculated from the best-fit model in the visibility analyses (\S~\ref{sec:mcmc}) is 14.9~$\pm$~0.9~K assuming a 10~\% uncertainty in the absolute flux calibration. 

We now estimate the physical properties of the dust ring. The optical depth $\tau_{\nu}$ is calculated as:
\begin{eqnarray}
I_{\nu} = \jh{B_{\nu}(T_{\rm mid})}\{1-{\exp}(-\tau_{\nu})\},
\end{eqnarray}
where $I_{\nu}$, $B_{\nu}$, and $T_{\rm mid}$ are the intensity, the Planck function, and the midplane temperature, respectively. We use the midplane temperature profile with a simplified expression for a passively heated, flared disk in radiative equilibrium \citep[e.g.,][]{dull2001}:
\begin{eqnarray}
T_{\rm mid}(r) = \left(\frac{\phi L_{*}}{8\pi r^{2}\sigma_{\rm SB}}\right)^{0.25},
\label{eq:tmid}
\end{eqnarray}
where $L_{*}$ is the stellar luminosity (0.017 to 0.113~$L_{\odot}$), $\phi$ is the flaring angle, and $\sigma_{\rm SB}$ is the Stefan--Boltzmann constant. If the starlight-absorbing small dust grains (sub-micron-sized) have a similar spatial distribution to the large dust grains (millimeter-sized) traced by millimeter emission (similar to the PDS~70 disk; \citealt{hashimoto2012, kepp19a}), the inner wall of the ring is directly exposed to stellar radiation, and $\phi$ reaches unity. In this case, at $r=$ 45~au where the asymmetry is located (see visibility analyses in \S~\ref{sec:mcmc}), Equation~(2) yields $T_{\rm mid} = 17.8$--$28.6$~K (uncertainty due to the stellar luminosity). The corresponding optical depth $\tau_{\nu}$ is 0.5--1.4 based on Equation~(1). Therefore, the asymmetry is marginally optically thin. The rest of the ring is expected to be optically thinner. Approximating the entire disk as an optically thin structure, the total flux corresponds to a total mass (gas~$+$~dust) of 7.6--15.7~$M_{\rm Jup}$ assuming \jh{the dust temperature of $T_{\rm mid} = 17.8$--$28.6$~K}, an opacity per unit dust mass of $\kappa_{\nu}=$3.45~cm$^2$~g$^{-1}$ at 345~GHz \citep{beckwith+91}, and a gas-to-dust mass ratio of 100. We note that the dust mass estimate is a lower limit \jh{since the actual dust temperature can be lower. The large dust grains are expected to concentrate at the pressure peak outside the cavity wall and reach a temperature lower than that estimated using Equation~\ref{eq:tmid} with $\phi\sim$unity. In the extreme case, the observed cavity in large dust grains may even be filled with small dust grains (e.g., ``missing cavities''; \citealt{dong12}), resulting in substantially lower temperature at the ALMA ring location. }

\subsection{Gas emission}

\begin{figure}[ht!]
\begin{centering}
\includegraphics[clip,width=\linewidth]{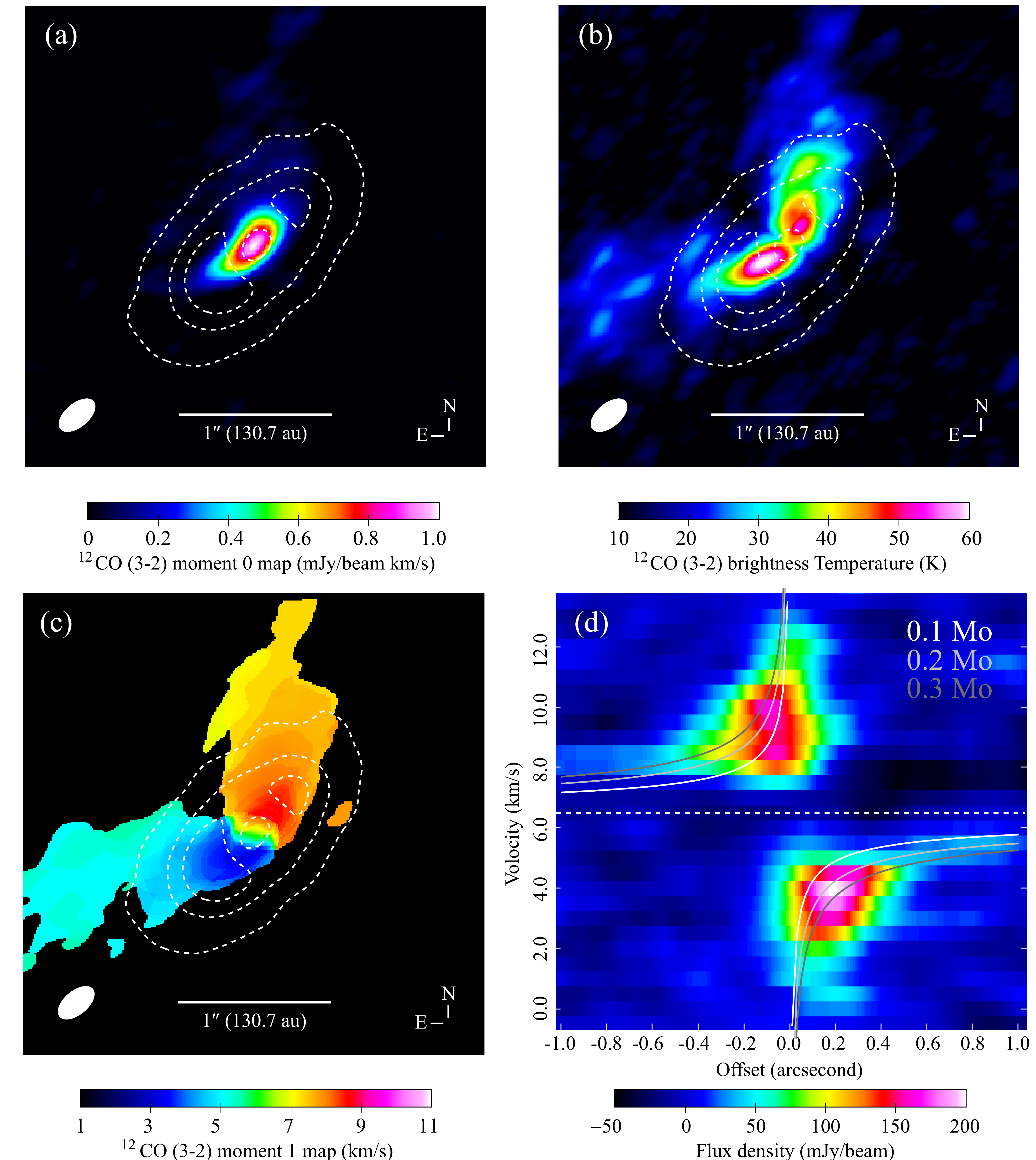}
\end{centering}
\caption{Synthesized images of the $^{12}$CO~$J=3\rightarrow2$ emission line for the ZZ~Tau~IRS ring.
(a) $^{12}$CO moment~0 map. The r.m.s.\ noise is 61.4~mJy/beam$\cdot$km/s with a beam size of 283~$\times$~156~mas at a PA of $-$50.2$\degr$. 
(b) $^{12}$CO moment~8 map converted to the brightness temperature. 
(c) $^{12}$CO moment~1 map. The r.m.s.\ noise is 9.56~mJy/beam at 0.5~km/s bin. The dotted contours represent the dust continuum at 10, 30, and 50~$\sigma$. 
(d) PV diagram along a PA of 135$\degr$ taken from \S~\ref{sec:mcmc}. The three lines denote loci of the peak emission of a Keplerian disk with a disk inclination of 60$\degr$ around 0.1 to 0.3~$M_{\odot}$ stars at $d=$ 130.7~pc. The systemic velocity (white dotted line) is assumed to be 6.5~km/s. 
} \label{fig:12co32}
\end{figure}

The integrated line flux map of $^{12}$CO~$J=3\rightarrow2$ in Figure~\ref{fig:12co32}(a) shows a single-peak structure with a peak flux of 1.0~$\pm$~0.1~Jy/beam$\cdot$km/s at 16.3~$\sigma$ and a total flux above 3~$\sigma$ of 1.9~$\pm$~0.19~Jy$\cdot$km/s assuming a 10~\% uncertainty in the absolute flux calibration. As shown in the channel maps in Figure~\ref{figA:channel} in the Appendix~\ref{secA:channel}, the $^{12}$CO emission is absorbed at 6.5--7.0~km/s, and thus the total flux could be a lower limit. The brightness temperature map of $^{12}$CO converted from the moment~8 map (maximum values of the spectrum) is shown in Figure~\ref{fig:12co32}(b). Since $^{12}$CO~$J=3\rightarrow2$ could be generally optically thick, it could be a good tracer for the gas temperature in the emitting region. At the location of the asymmetry, the brightness temperature of $^{12}$CO is $\sim$35~K, higher than the 17.8--28.6~K estimated by Equation~(2). This could be because the emitting surface of $^{12}$CO is in the upper hotter layer of the disk. Figure~\ref{fig:12co32}(d) shows the $^{12}$CO position--velocity (PV) diagram (channel maps: Figure~\ref{figA:channel} in the Appendix~\ref{secA:channel}) along PA $=$ 135$\degr$. We overplotted  the loci of the peak emission of a Keplerian disk with an inclination of 60$\degr$ (see \S~\ref{sec:mcmc}) around a central star with masses of 0.1, 0.2, and 0.3~$M_{\odot}$, and found that the dynamical mass of ZZ~Tau~IRS is $\sim$0.1--0.3~$M_{\sun}$. This value is consistent with the spectroscopically determined mass (0.1--0.2~$M_{\sun}$; \citealp{andr2013,herczeg2014}). Note that we assume that the systemic velocity of ZZ~Tau~IRS is 6.5~km/s in the PV diagram.

\section{Visibility analyses} \label{sec:mcmc}

Since the contrast of the asymmetry in the ZZ~Tau~IRS ring is of order unity, an imaging artifact such as a sidelobe of the bright ring could generate an apparent asymmetry. To characterize the disk structure, we performed forward modelling in which the observed visibilities are reproduced with a parametric model of the ring in the visibility domain. 

\begin{figure}[ht!]
\begin{centering}
\includegraphics[clip,width=\linewidth]{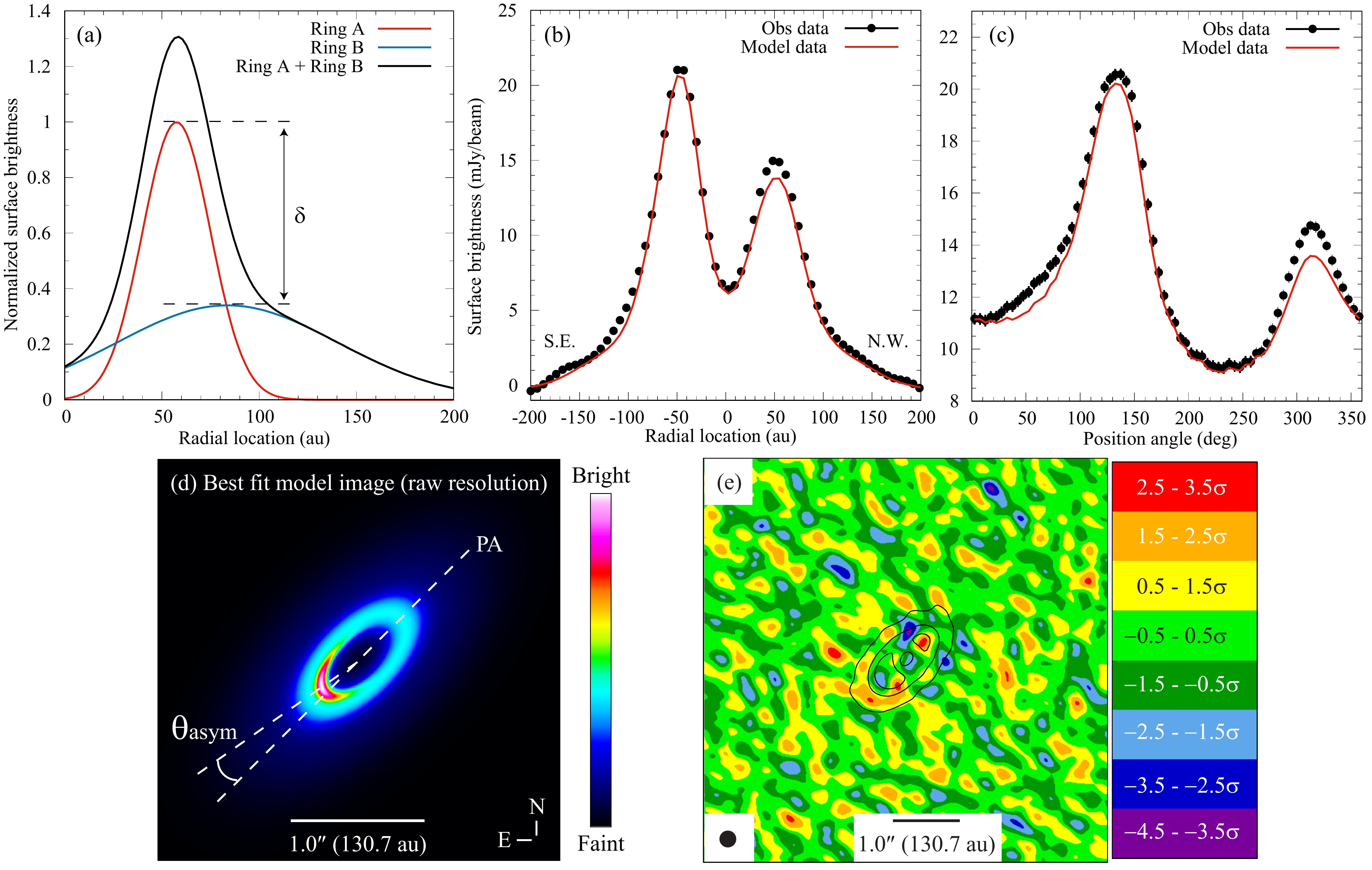}
\end{centering}
\caption{(a) Azimuthally averaged radial surface brightness for the best-fit model \jh{with a raw resolution}. The asymmetry is not included in this panel. \jh{(b and c) Radial and azimuthal profiles of the observations and the best-fit model convolved with the beam in observations of 254~$\times$~248~mas at a PA of $-$61.7\degr. Observed data are same with those in Figure~\ref{fig:dustcontinuum}(b) and (c).} (d) Best-fit model image (3$''$~$\times$~3$''$). (e) Residual image (6$''$~$\times$~6$''$). The reduced-$\chi^{2}$ is 1.91. The black dotted contours represent the dust continuum image at 10, 30, and 50~$\sigma$. 
}\label{fig:mcmc}
\end{figure}

\jh{The ZZ~Tau~IRS disk is assumed to consist of a ring and an asymmetry. The model ring consists of a narrow Gaussian ring (ring~A) superposed on a wide Gaussian ring (ring~B) as shown in Figure~\ref{fig:mcmc}(a), and is described as follows:}
\begin{eqnarray}
I_{\rm ring}(r) \propto  
{\rm exp}\left(-\frac{(r - r_{\rm peak,ringA})^{2}}{2\sigma^{2}_{r,{\rm ringA}}} \right) + 
\delta\ {\rm exp}\left(-\frac{(r - r_{\rm peak,ringB})^{2}}{2\sigma^{2}_{r,{\rm ringB}}} \right),
\end{eqnarray}
where $r_{\rm peak,ringA}$, $\sigma_{r,\rm{ringA}}$, $r_{\rm peak,ringB}$, $\sigma_{r,\rm{ringB}}$, and $\delta$ are the radial peak positions and standard deviations of two Gaussian rings and a scaling factor between the two, respectively. The width of the ring (FWHM) is calculated as 2.355$\sigma_{\rm ring}$.

The asymmetry is defined as an elliptical Gaussian function in polar coordinates as follows:

\begin{eqnarray}
I_{\rm asym}(r,\theta) \propto
{\rm exp} \left(-\frac{(r - r_{\rm peak,asym})^{2}}{2\sigma^{2}_{r{\rm,asym}}}-\frac{(\theta - \theta_{\rm peak,asym})^{2}}{2\sigma^{2}_{\theta{\rm,asym}}}\right), 
\end{eqnarray}
where $r_{\rm peak,asym}$, $\theta_{\rm peak,asym}$, $\sigma_{r\rm{,asym}}$, and $\sigma_{\theta{\rm,asym}}$ are the radial peak position, azimuthal peak position, radial standard deviation, and azimuthal standard deviation, respectively. The flux for the asymmetry is normalized to $F_{\rm asym}$. The combined model image is magnified and rotated with an inclination ($i$) and a PA. The total flux for the combined model image is normalized to $F_{\rm total}$. The radial and azimuthal widths (FWHM) of the asymmetry are 2.355$\sigma_{r,\rm{asym}}$ and 2.355$\sigma_{\theta,\rm{asym}}$, respectively. In total, there are 13 free parameters in our model ($F_{\rm total}$, $F_{\rm asym}$, $r_{\rm peak,ringA}$, $\sigma_{r,\rm{ringA}}$, $r_{\rm peak,ringB}$, $\sigma_{r\rm{,ringB}}$, $\delta$, $r_{\rm peak,asym}$, $\theta_{\rm peak,asym}$, $\sigma_{r,\rm{asym}}$, $\sigma_{\theta{\rm,asym}}$, $i$, PA).

The model image was converted to complex visibilities with the public Python code \verb#vis_sample# \citep{loomis+17}, in which model visibilities are samples with the same ($u$, $v$) grid points as the observations. The model visibilities are deprojected\footnote{Visibilities are deprojected in the $uv$-plane using the following equations \citep[e.g.,][]{zhan16}: 
 $u'    =   (u\,{\rm cos}\,{\rm PA} - v\,{\rm sin}\,{\rm PA}) \times {\rm cos}\,i,
  v'    =   (u\,{\rm sin}\,{\rm PA} - v\,{\rm cos}\,{\rm PA}),$
where $i$ and PA are free parameters in our visibility analyses in \S~\ref{sec:mcmc}.} with the system PA and $i$ as free parameters. The fitting is performed with a Markov chain Monte Carlo (MCMC) method in the \verb#emcee# package \citep{foreman-mackey+2013}. The log-likelihood function ln$L$ in visibility analyses is 
\begin{eqnarray}
{\rm ln}L = -0.5 \sum \left[W_{j}\{({\rm Re}V_{j}^{\rm obs} - {\rm Re}V_{j}^{\rm model})^{2} + ({\rm Im}V_{j}^{\rm obs} - {\rm Im}V_{j}^{\rm model})^{2}\}\right],
\end{eqnarray}
where the subscript $j$ represents the $j$-th data. $V_{j}^{\rm obs}$, $V_{j}^{\rm model}$, and $W_{j}$ are observed and model visibilities and weights corresponding to 1/$\sigma_{j}^{2}$ \jh{($\sigma_j$ is the rms noise of a given visibility. See details in CASA Guides\footnote{{\sf https://casaguides.nrao.edu/index.php/DataWeightsAndCombination}}.), respectively. The values of $V_{j}^{\rm obs}$ and $W_{j}$  (i.e., 1/$\sigma_{j}^{2}$) are provided in the measurement set of ALMA data.} Our calculations used flat priors with the parameter ranges summarized in Table~\ref{tab:mcmc}. We ran 3,000 steps with 100~walkers, and discarded the initial 1,000~steps as the burn-in phase based on trace plots in Figure~\ref{figA:trace2} in Appendix~\ref{secA:trace2}.

The fitting results with uncertainties computed from the 16th and 84th percentiles, the radial profile of best-fit surface brightness for the ring \jh{with a raw resolution}, the best-fit model image,  and the probability distributions for the MCMC posteriors are shown in Table~\ref{tab:mcmc}, Figure~\ref{fig:mcmc}(a), Figure~\ref{fig:mcmc}(d), and Figure~\ref{figA:corner2} in the Appendix~\ref{secA:corner2}, respectively. We subtracted modeled visibilities from observed visibilities \jh{(Figure~\ref{figA:vis} in Appendix~\ref{secA:vis})}, and made a CLEANed image (Figure~\ref{fig:mcmc}e) in which we did not find significant residuals. To calculate the value of the reduced-$\chi^{2}$, we derived a factor between the weights and standard deviations \jh{(referred to as stddev)} in the visibilities by calculating the standard deviations of the real and imaginary parts in 2~k$\lambda$ bins along the azimuthal direction in the visibility domain. \jh{The values of weights $W_j$ are provided in the measurement set of ALMA data while the standard deviations (stddev) are calculated in our fitting.} The visibilities were deprojected with $i=$ 60$\degr$ and PA~$=$ 135$\degr$. Figure~\ref{figA:weight} in Appendix~\ref{secA:weight} shows a comparison between the weights and standard deviations (stddev$^{-2}$). We found that the weights are typically 4.0$\times$ higher than the standard deviations. Finally, we calculated a reduced-$\chi^{2}$ of 1.91. We also conducted visibility analyses using a model with one Gaussian ring $+$ one asymmetry (i.e., $\delta=$ 0), and found significant residuals. Thus, the model with two Gaussian rings $+$ one asymmetry is necessary to reproduce the observed dust continuum image.

\begin{splitdeluxetable*}{cccccccBcccccc}
\tablewidth{0pt} 
\tablenum{2}
\tablecaption{Results of visibility analyses and parameter ranges \label{tab:mcmc}}
\tablehead{
\colhead{$r_{\rm peak,ringA}$} & \colhead{FWHM$_{\rm{ringA}}$} & \colhead{$r_{\rm peak,ringB}$} & \colhead{FWHM$_{\rm{ringB}}$} & \colhead{$\delta$} & \colhead{$i$} & \colhead{PA} & \colhead{$F_{\rm total}$} & 
\colhead{$F_{\rm asym}$} & \colhead{$r_{\rm peak,asym}$} & \colhead{$\theta_{\rm peak,asym}$} & \colhead{FWHM$_{r\rm{,asym}}$} & \colhead{FWHM$_{\theta{\rm,asym}}$} \\
\colhead{(au)} & \colhead{(au)}  & \colhead{(au)} & \colhead{(au)} & \colhead{} & \colhead{(deg)}           & \colhead{(deg)} & 
\colhead{(mJy)} & \colhead{(mJy)}       & \colhead{(au)}       & \colhead{(deg)} & \colhead{(au)}   & \colhead{(deg)}                
}
\colnumbers
\startdata
57.51$^{+0.19}_{-0.18}$ & 41.08$^{+0.63}_{-0.60}$ & 83.62$^{+1.61}_{-1.67}$ & 133.65$^{+1.59}_{-1.62}$ & 0.34$^{+0.01}_{-0.01}$ & 60.16$^{+0.07}_{-0.07}$ & 134.73$^{+0.09}_{-0.09}$ & 
273.94$^{+0.66}_{-0.67}$ & 14.51$^{+0.25}_{-0.24}$ & 45.15$^{+0.22}_{-0.21}$ & $-$18.17$^{+0.52}_{-0.52}$ & 11.31$^{+0.98}_{-1.01}$  & 103.07$^{+1.59}_{-1.60}$ \\ 
\{0.0 .. 117.6\} & \{0.0 .. 78.4\} & \{0.0 .. 117.6\} & \{78.4 .. 196.1\} & \{0.0 .. 1.0\} & \{50.0 .. 70.0\} & \{125.0 .. 145.0\} & 
\{250.0 .. 300.0\} & \{0.0 .. 100.0\} & \{0.0 .. 117.6\} & \{$-$45.0 .. $+$45.0\} & \{0.0 .. 117.6\} & \{0.0 .. 360.0\} \\
\enddata
\tablecomments{
 Parentheses describe parameter ranges in our MCMC calculations. }
\end{splitdeluxetable*}

Though the observed dust continuum image in Figure~\ref{fig:dustcontinuum}(a) shows two peaks along the ring major axis, our visibility analyses suggest that only the one at PA $=$ 130$\degr$ is real.

The radial ring position (i.e., the peak position) and its FWHM (Ring~A $+$ Ring~B) in the model image were measured to be 58 and 49~au in Figure~\ref{fig:mcmc}(a). The contrast in the asymmetry, which is the ratio between the peak brightness at the asymmetry and that at its opposite side on the ring, is measured to be 3.0 in Figure~\ref{fig:mcmc}(d). 

\section{Discussion} \label{sec:discuss}

\subsection{Azimuthal asymmetry}
\begin{figure}[ht!]
\centering
\includegraphics[clip,width=170mm]{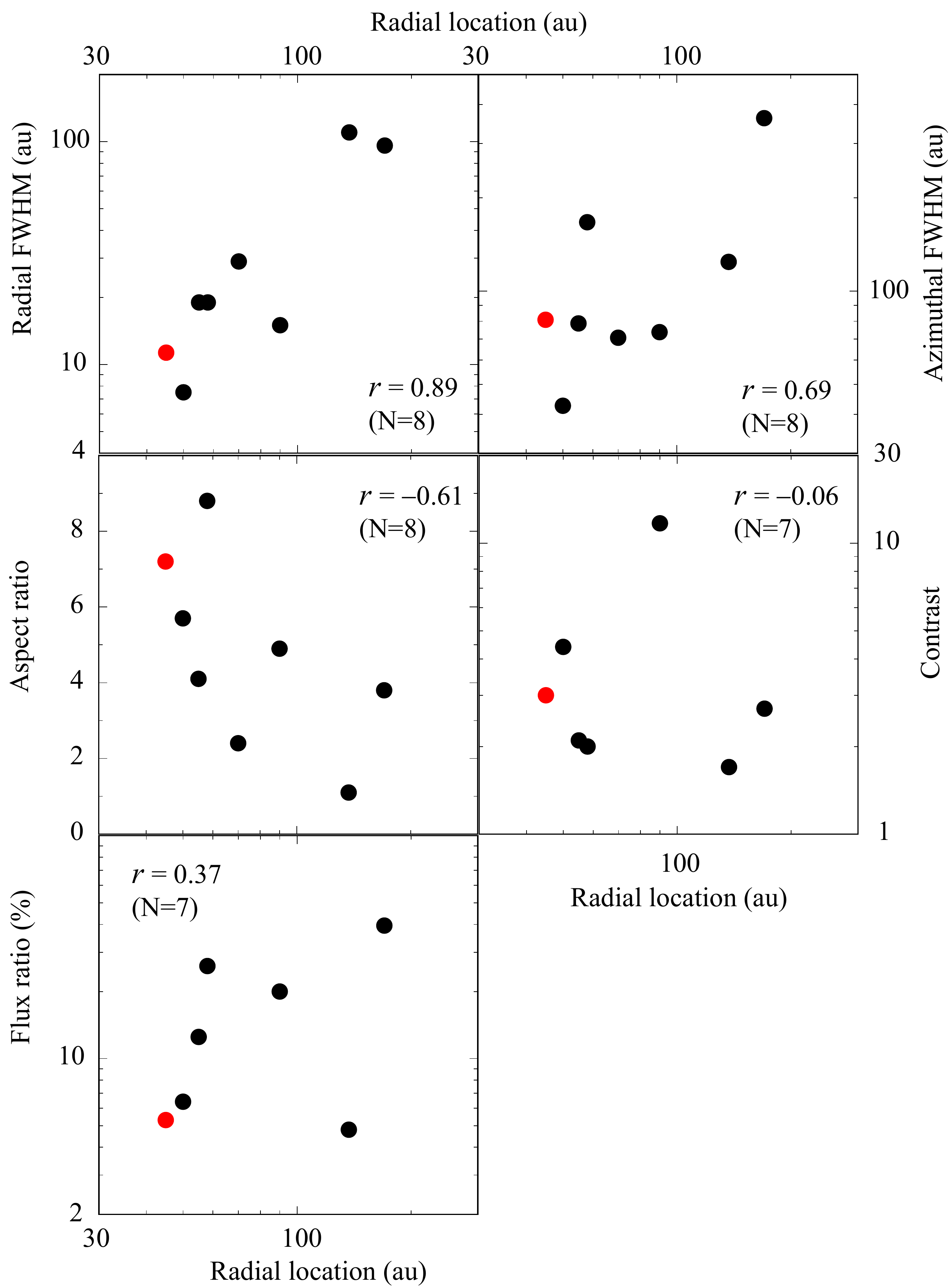}
\caption{Comparison of the properties of the ZZ~Tau~IRS asymmetry with other asymmetries around intermediate mass stars as a function of the radial location. The red circle represents ZZ~Tau~IRS. The physical quantities and their brief explanations are summarized in Table~\ref{tab:comparison}. \jh{Pearson's correlation coefficient $r$ is inserted in panels. The 5\% significance values for Pearson's $r$ with $n=$7 and 8 are 0.75 and 0.71, respectively \citep{pugh1966}.}
}\label{fig:comparison}
\end{figure}

Azimuthal asymmetries in rings have been identified in roughly ten protoplanetary disks \citep[e.g.,][]{vandermarel20a}. At band 7, their shape parameters (e.g., the radial FWHM and the aspect ratio defined as the azimuthal FWHM divided by the radial FWHM) are summarized in Table~\ref{tab:comparison}\footnote{The physical quantities of other asymmetries are taken from \citet{vandermarel20a}. V1247~Ori shows a crescent structure in ALMA band~7 \citep{kraus+17}. However, we do not include V1247~Ori because the structure does not have a Gaussian profile \citep{kraus+17}. Though HD~142527 also shows a crescent structure in ALMA band~7 \citep{fukagawa+13}, the visibility analyses in \citet{vandermarel20a} did not find convergence; thus, we do not include HD~142527.}. All but ZZ~Tau~IRS are intermediate mass stars ($\sim$2~$M_{\sun}$). In Figure~\ref{fig:comparison} we compare various properties of these asymmetries as a function of the peak radial location. Though ZZ~Tau~IRS is a VLM star, its asymmetry has similar properties to other asymmetries around more massive stars, implying a similar origin.  

\jh{We explore the correlation in Figure~\ref{fig:comparison} by calculating Pearson's correlation coefficient $r$ for logarithmic numbers except the aspect ratio. The 5\% significance values for Pearson's $r$ with $n=$7 and 8 are 0.75 and 0.71, respectively \citep{pugh1966}. We found two (marginally) statistically significant relations: radial location {\it v.s.} radial FWHM ($r=$0.89), and radial location {\it v.s.} azimuthal FWHM ($r=$0.69; top panels in Figure~\ref{fig:comparison}). Asymmetric structures have been interpreted as dust traps at pressure maxima in gas vortices \citep[e.g.,][]{raettig2015} or horseshoes \citep[e.g.,][]{ragusa2017}. The radial width of a gas pressure bump is expected to be comparable to the gas scale height \citep[e.g.,][]{dullemond2018a}, typically prescribed as $H(r) \propto r^q$ with $q\sim$ unity (e.g., assuming $T(r) \propto r^{-0.5}$, $q=1.25$). The most significant correlation we see in Figure~\ref{fig:comparison}, radial location {\it v.s.} radial FWHM, may simply trace the dependence of the radial size of gas pressure structure on $r$.}

\jh{Another interpretation of asymmetries is that they} are part of spiral arms induced by companions \citep[e.g.,][]{vandermarel+16a}. One way to distinguish these mechanisms is to derive the spectral index from multi-wavelength observations, which may constrain the size of the dust grains \citep[e.g.,][]{test2014}. Dust trapping in gas vortices and horseshoes results in a bigger dust size, while spiral arms are not expected to trap dust particles as they deferentially rotate with respect to the local disk material \citep{papaloizou2007, li2020a}. Measuring the pattern speed of asymmetries may also shed light on their origin, as vortices and horseshoes are expected to move at the local Keplerian velocity while spirals corotate with their drivers.

\begin{deluxetable*}{ccccccccc}
\tabletypesize{\footnotesize}
\tablewidth{0pt} 
\tablenum{3}
\tablecaption{Stellar mass and physical quantities for asymmetries \label{tab:comparison}}
\tablehead{
\colhead{Object} & \colhead{Stellar mass} & \colhead{Radial location} & \colhead{Radial FWHM} & \colhead{Azimuthal FWHM} & \colhead{Aspect ratio} & \colhead{Contrast} & \colhead{Flux ratio} & \colhead{Refs} \\
\colhead{} & \colhead{($M_{\sun}$)} & \colhead{(au)}        & \colhead{(au)}         & \colhead{(deg; au)}    &\colhead{}      &\colhead{}      &\colhead{(\%)}      & \colhead{} 
}
\colnumbers
\startdata
ZZ Tau IRS & 0.2 & 45  & 11.3 & 103; 80.9  & 7.2 & 3.0  & 5.3  & a,a \\
MWC 758    & 1.9 & 50  & 7.5  & 49; 42.7   & 5.7 & 4.4  & 6.4  & b,c \\
           &     & 90  & 15   & 47; 73.8   & 4.9 & 11.7 & 20.0 & b,c \\
HD 34282   & 2.0 & 137 & 110  & 52; 124.3  & 1.1 & 1.7  & 4.8  & b,d \\
SR 21      & 2.1 & 55  & 19   & 82; 78.7   & 4.1 & 2.1 & 12.5 & b,c \\
           &     & 58  & 19   & 165; 166.9 & 8.8 & 2.0 & 26.0 & b,c \\
AB Aur     & 2.3 & 170 & 96   & 122; 361.8 & 3.8 & 2.7 & 39.5 & b,c \\
IRS 48     & 2.2 & 70  & 29   & 58; 70.8   & 2.4 & --- & ---  & b,c \\
\enddata
\tablecomments{
(1) Object name. MWC~758 and SR~21 have two asymmetries \citep{vandermarel20a}. (2) Stellar mass. (3) Radial location. (4) Radial FWHM calculated by 2.355$\sigma_{r}$. The radial FWHM of the IRS~48 asymmetry is calculated with 2.17$\sigma_{r}$ because the radial profile for IRS~48 was found to be best fit with a 4th power in 2D Gaussian in \citet{vandermarel20a}. (5) Azimuthal FWHM. (6) The aspect ratio defined as the azimuthal FWHM divided by the radial FWHM. (7) Contrast measured in the model between the peak brightness and the opposite side of the ring. The model images are constructed with the parameters from the literature in column (9). For SR~21, since the azimuthal positions of the two asymmetries are located at almost opposite sides, the contrast of the each asymmetry was measured for a model containing only one asymmetry and the ring. For IRS~48, the contrast is not measured because the model contains only the asymmetry without the ring \citep{vandermarel20a}. (8) Ratio of the asymmetry to total flux in the model. For IRS~48, this value is not measured because the model contains only the asymmetry without the ring \citep{vandermarel20a}. (9) References for information of stellar masses and properties of asymmetries: a) this work, b) \citet{garu2018}, c) \citet{vandermarel20a}, d) \citet{vanderpalas17}. }
\end{deluxetable*}

\subsection{Gravitational stability}

\jh{ZZ~Tau~IRS' disk is one of the most massive among VLM stars \citep[cf.,][]{Kurtovic2020a}. We examine its gravitational stability. Theoretical studies \citep[e.g.,][]{durisen2007} suggest that if the \citet{toomre1964} Q-parameter, defined as  
\begin{eqnarray}
Q=\frac{c_{\rm s} \Omega_{\rm K}}{\pi G \Sigma_{\rm disk}},
\end{eqnarray}
where $c_{\rm s}$, $\Omega_{\rm K}$, and $\Sigma_{\rm disk}$ are the sound speed, the Keplerian angular velocity, and the disk (gas) surface density, respectively, is of order of unity, the disk may be subject to gravitational instability (GI). With the scale height of $H \sim c_{\rm s} / \Omega_{\rm K}$, Toomre's $Q$-parameter can be described as $Q \sim H / r \times M_{\rm star} / M_{\rm disk}$, where $M_{\rm disk} = \pi r^{2} \Sigma_{\rm disk}$. Therefore, if the disk's aspect ratio is $H / r \sim 0.1$,  GI occurs in massive disks with  $M_{\rm disk}/M_{\rm star} \gtrsim$0.1.}
The disk-to-star mass ratio is $\sim$0.04--0.08 in ZZ~Tau~IRS (\S~\ref{subsec:dust}), and thus the disk may be globally gravitationally stable. Meanwhile the Q-value at the asymmetry is \jh{1.7--6.2 assuming a local surface density $\Sigma_{\rm disk} = {\rm cos}\,i\,\tau_\nu$/$\kappa_\nu=$ 7--20~g/cm$^{2}$} ($i=60\degr$ and $\tau_\nu=$0.5--1.4, \S~\ref{subsec:dust}), an opacity per unit dust mass of $\kappa_{\nu}=$ 3.45~cm$^2$~g$^{-1}$ at 345~GHz \citep{beckwith+91}, a temperature of 17.8--28.6~K (\S~\ref{subsec:dust}), and a gas-to-dust mass ratio of 100. \jh{Hence, the asymmetry may also be marginally stable.} 

\subsection{Possible warped inner disk}

ZZ~Tau~IRS has been suggestsed to have an edge-on inner disk \citep{white2004,furlan2011}, because the IR excess is larger than the \jh{optical} stellar flux \jh{due to partial disk obscuration of the star} (Figure~\ref{figA:sed} in the Appendix~\ref{secA:sed}). Our visibility analyses suggest that the outer ring  at $r=$ 50~au has an inclination ($i$) of 60$\degr$ (\S~\ref{sec:mcmc}). Radiative transfer calculations for protoplanetary disks (full disks) show that typically the central star starts to be significantly obscured by the disk once the inclination is larger than $\sim$70$\degr$ \citep[e.g., see Figure~2 in][]{whit2013}. Hence ZZ~Tau~IRS could possess an inner disk misaligned with the outer ring. Such warped inner disks have been discovered in twisted gas flow patterns \citep[e.g.,][]{rose14,mayama2018} and shadows on the outer disk in NIR scattering images \citep[e.g.,][]{mari2015}. For ZZ~Tau~IRS, current ALMA gas observations in Figure~\ref{fig:12co32} are insufficient to confirm a twisted gas flow due to low spatial \& spectral resolution. Furthermore, high spatial resolution observations of NIR scattering images have not yet been carried out. Another possible explanation of the observed SED is that the source is a ``flat spectrum'' young stellar object surrounded by an infalling dusty envelope \citep{calvet1994}. In this case, an extended reflection nebulae around the source is expected in NIR scattered light observations (e.g., similar to T~Tau, \citealp{whitney1993a}). Future observations are necessary to test these hypotheses.

A few possible mechanisms have been proposed to explain warped disks. An exciting one is disk--companion interactions \citep[e.g.,][]{zhu2019a}, in which a massive companion with a companion-star mass ratio $q\gtrsim0.01$ (or $M_{\rm p}\gtrsim M_{\rm J}$ around a 0.1$M_\odot$ star) on an inclined orbit induces both a deep wide gap and a misaligned inner disk. Such a system with a massive companion and a misaligned inner disk has been reported in HD~142527 \citep{bill2012,mari2015}. Future detection of massive companions in the gap would serve as a test for the planet origin hypothesis.

\section{Conclusion} \label{sec:conclusion}

We analyzed ALMA band~7 archive data (dust continuum and $^{12}$CO~$J=3\rightarrow2$ emission) of ZZ~Tau~IRS. The source has been classified as a VLM star with a mass of $\sim$0.1--0.2~$M_{\sun}$ in the literature. The observed $^{12}$CO~$J=3\rightarrow2$ kinematics suggest a dynamical mass of $\sim$0.1--0.3~$M_{\sun}$, consistent with previous estimates. The ALMA dust continuum image at a spatial resolution of 0\farcs25 shows a ring and an asymmetric structures around ZZ~Tau~IRS. Our visibility analysis confirmed that the observed asymmetry is real rather than apparent.

Our best-fit model of the ALMA dust observations shows that the asymmetry has a radial FWHM of 11.3~au, an aspect ratio of 7.2 (ratio of azimuthal FWHM to azimuthal FWHM),  a brightness contrast of 3.0, and the asymmetry contributes 5\% of the total disk flux. These metrics are comparable with asymmetries seen around intermediate mass stars ($\sim$2~$M_{\sun}$), implying a similar origin, despite ZZ~Tau~IRS being 10 times less massive. While future high spatial resolution observations in multiple wavelengths with ALMA and JVLA are needed to determine the exact origin of the asymmetry, the three leading hypotheses all invoke companions inside the gap.

The inner disk around ZZ~Tau~IRS, not visible in our observations, has been inferred to be edge-on based on the SED. Our observations showed that the inclination of the outer disk at $r=$ 50~au is 60\degr. These results imply that the ZZ~Tau~IRS inner disk may be misaligned relative to the outer ring, and it is possible that the misalignment (and the gap) is produced by a giant planet with $M_{\rm p}\gtrsim M_{\rm J}$ on an inclined orbit inside the gap. Future scattered light observations may test this hypothesis by searching for shadows on the outer disk cast by the inner disk. 

Overall, the structures found in the ZZ Tau IRS disk suggest the presence of \jh{massive companions including gas giant planets} at $\sim10$~au. This is surprising as the core accretion theories have generally prohibited the formation of such planets around VLM stars (\citealp{miguel2020a,liu2020a}, \S~\ref{sec:intro}). Hence, ZZ~Tau~IRS will be a prime target in the investigation of planet formation around VLM stars.

\acknowledgments
\jh{We thanks the anonymous referee for a helpful review of the manuscript.
We also thanks Beibei Liu for discussions.} 
This work was supported by JSPS KAKENHI Grant Numbers 19H00703, 19H05089, and 19K03932, 18H05441, 17H01103. R.D. acknowledges support from the Alfred P. Sloan Foundation and the Natural Sciences and Engineering Research Council of Canada.
This paper makes use of the following ALMA data: ADS/JAO.ALMA\#2016.1.01511.S. ALMA is a partnership of ESO (representing its member states), NSF (USA), and NINS (Japan), together with NRC (Canada), NSC (Taiwan), ASIAA (Taiwan), and KASI (Republic of Korea), in cooperation with the Republic of Chile. The Joint ALMA Observatory is operated by ESO, AUI/NRAO, and NAOJ.
\jh{This publication makes use of data products from the Two Micron All Sky Survey, which is a joint project of the University of Massachusetts and the Infrared Processing and Analysis Center/California Institute of Technology, funded by the National Aeronautics and Space Administration and the National Science Foundation.
This publication makes use of data products from the Wide-field Infrared Survey Explorer, which is a joint project of the University of California, Los Angeles, and the Jet Propulsion Laboratory/California Institute of Technology, funded by the National Aeronautics and Space Administration.}

{\it Software}: \verb#vis_sample# \citep{loomis+17}, 
          \verb#CASA# \citep{mcmu07}, 
          \verb#emcee# \citep{foreman-mackey+2013}

\appendix

\section{SED of ZZ Tau IRS}\label{secA:sed}

\begin{figure}[ht!]
\begin{centering}
\includegraphics[clip,width=\linewidth]{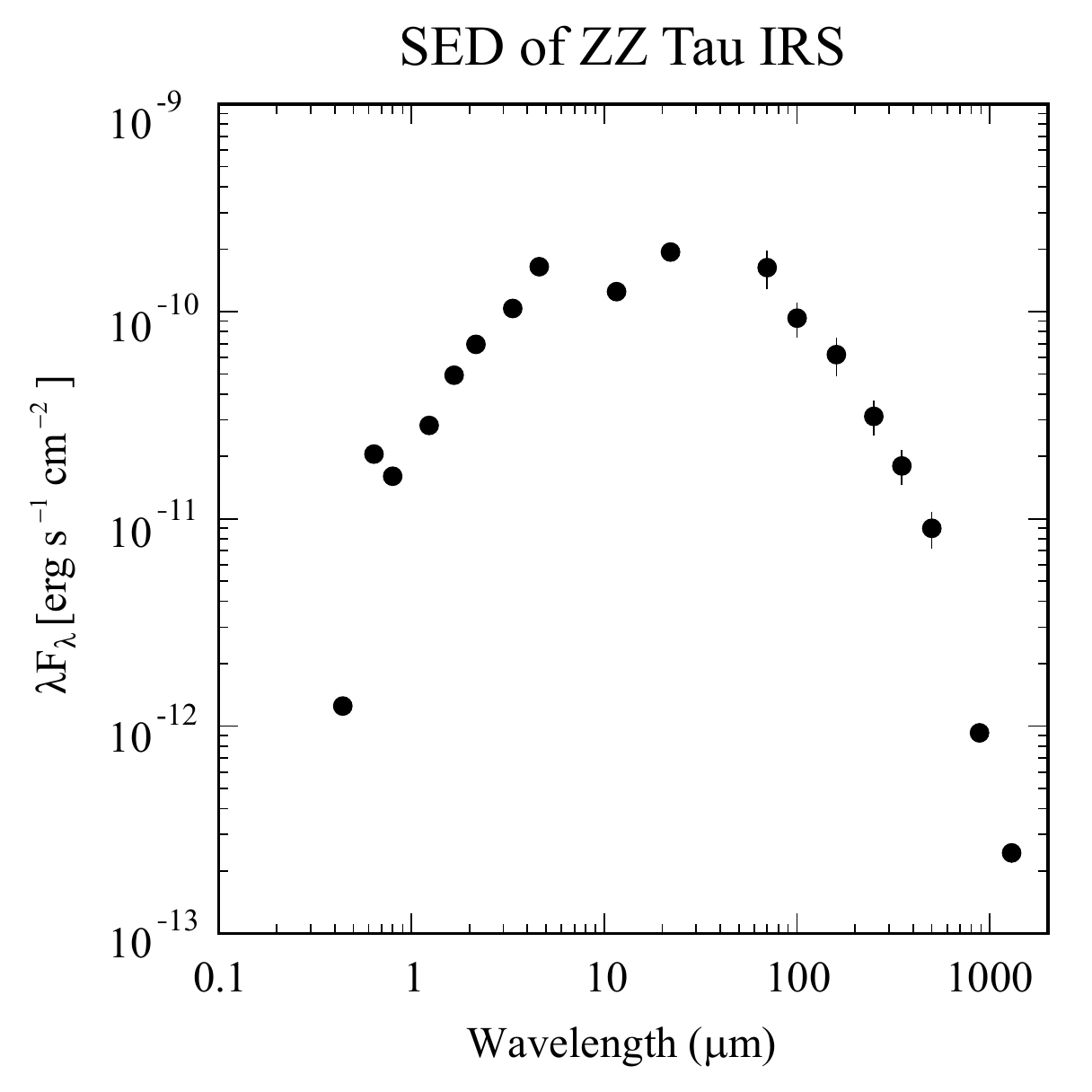}
\end{centering}
\caption{SED of ZZ Tau IRS without extinction correction. \jh{The photometric data are summarized in Table~\ref{tab:phot}.}}\label{figA:sed}
\end{figure}

\jh{Figure}~\ref{figA:sed} shows the SED of ZZ Tau IRS. \jh{Extinction correction is not applied to show the large IR excess relative to the stellar flux. The photometric data without extinction correction are summarized in Table~\ref{tab:phot}.}

\begin{deluxetable*}{lll}
%\tabletypesize{\scriptsize}
\tablewidth{0pt} 
\tablenum{4}
\tablecaption{Photometric data used in the SED of ZZ~Tau~IRS\label{tab:phot}}
\tablehead{
\colhead{Wavelength} & \colhead{$\lambda F_{\lambda}$} & \colhead{Reference}  \\
\colhead{($\mu$m)}   & \colhead{(10$^{-12}$ erg s$^{-1}$ cm$^{-2}$)} & \colhead{}  \\
}
%\colnumbers
\startdata
0.44         & 1.25                 & USNO-B1.0 \citep{mone2003} \\
0.64         & 20.5                 & USNO-B1.0 \citep{mone2003} \\
0.80         & 16.1                & USNO-B1.0 \citep{mone2003} \\
1.23        & 28.3  $\pm$ 0.6     & 2MASS \citep{cutr2003a} \\
1.66        & 49.2  $\pm$ 1.0     & 2MASS \citep{cutr2003a} \\
2.16        & 69.4  $\pm$ 1.5     & 2MASS \citep{cutr2003a}\\
3.35        & 100.4 $\pm$ 2.0       & WISE  \citep{cutr2014a}\\
4.60        & 164.5 $\pm$ 3.2       & WISE  \citep{cutr2014a} \\
11.6        & 124.8 $\pm$ 1.6       & WISE  \citep{cutr2014a} \\
22.1        & 193.8 $\pm$ 2.9       & WISE  \citep{cutr2014a} \\
70          & 162.9 $\pm$ 34.3 & Herschel \citep{ribas2017a}\\
100         & 93.0  $\pm$ 18.0 & Herschel \citep{ribas2017a}\\
160         & 61.9  $\pm$ 13.1 & Herschel \citep{ribas2017a}\\
250         & 31.2  $\pm$ 6.0 & Herschel \citep{ribas2017a}\\
350         & 18.0  $\pm$ 3.4 & Herschel \citep{ribas2017a}\\
500         & 9.0   $\pm$ 1.8 & Herschel \citep{ribas2017a}\\
885         & 0.928 $\pm$ 0.093  & ALMA (this work)\\
1300        & 0.244 $\pm$ 0.003 & SMA \citep{andr2013} \\
\enddata
\tablecomments{
Extinction correction is not applied here.}
\end{deluxetable*}

\section{Dust continuum images synthesized with only the imaginary part}\label{secA:imag}

\begin{figure}[ht!]
\begin{centering}
\includegraphics[clip,width=\linewidth]{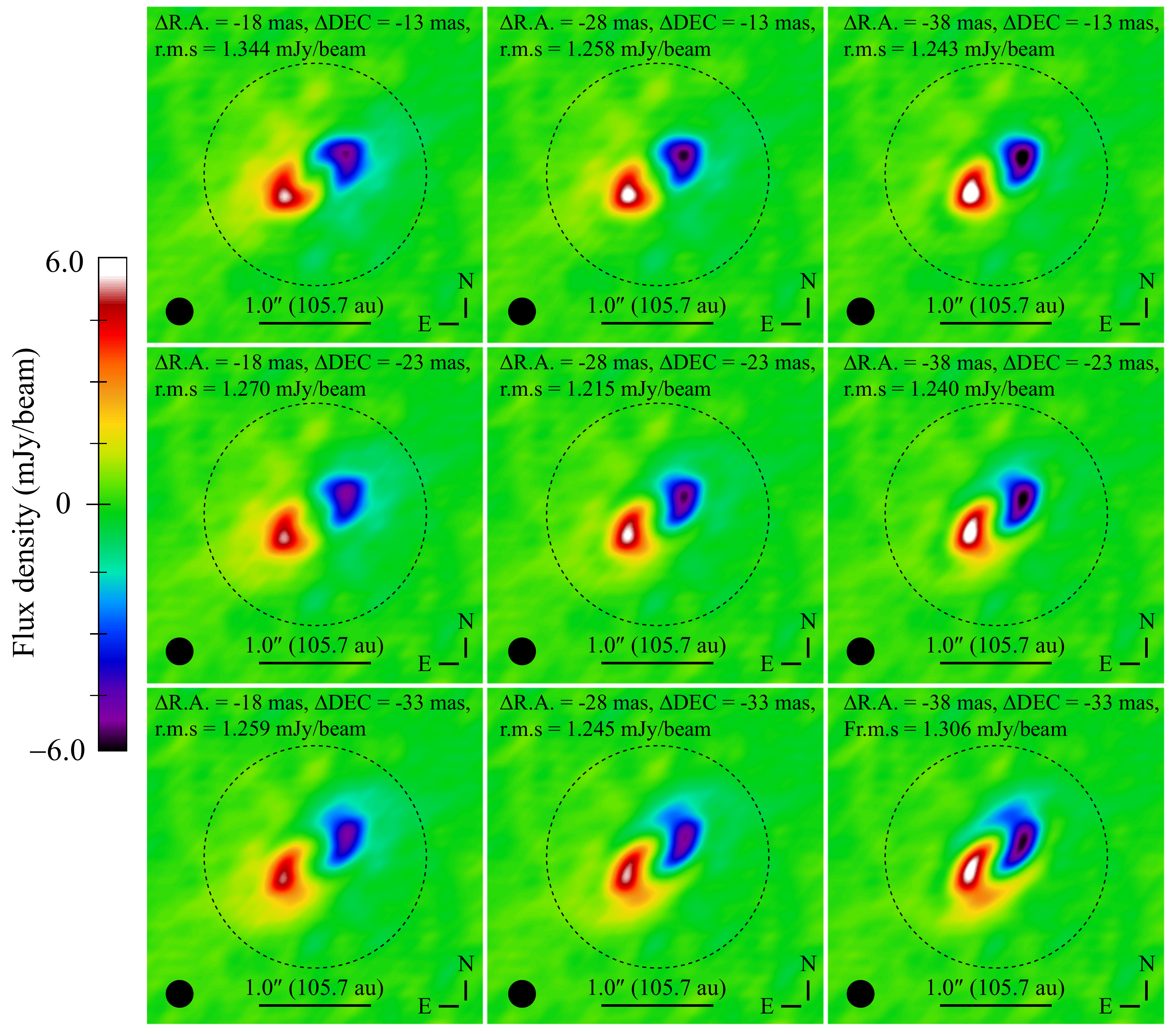}
\end{centering}
\caption{Dust continuum images synthesized with only the imaginary part of the long baseline data by shifting by 10~mas in the R.A. and DEC directions. The r.m.s.\ value is measured inside the black dotted circle with a diameter of 2\farcs25. The 1-$\sigma$ noise is 0.268~mJy/beam measured in a region far from the object. The beam size is 254~$\times$~248~mas at a PA of $-$61.7\degr.}\label{figA:long}
\end{figure}

We shifted the observed dust continuum images to minimize the asymmetry relative to the phase center as follows. First, the proper motions of ZZ~Tau~IRS were calculated with the function \verb#EPOCH_PROP# in GAIA ADQL (Astronomical Data Query Language\footnote{{\sf https://gea.esac.esa.int/archive/}}). The phase centers were corrected to (4h30m51.7346s, $+$24d41m47.175s) in ICRS using \verb#fixvis# in the CASA tools. Next, in order to make the analyses of the disk structure more effective, we further shift the center of the image by searching for the location where the disk asymmetry is minimized. We subtracted the 180\degr-rotated image in the visibility domain. This procedure corresponds to producing a synthesized image with only the imaginary part of the visibilities. 
\jh{This method has been utilized in the analyses of DM~Tau \citep{Hashimoto2021a} and WW~Cha \citep{Kanagawa2021a}. Since the visibility $V$ is a Fourier transform of the surface brightness distribution of the sky, which is a real quantity, the visibility at $(-u, -v)$ in the $uv$-plane is a complex conjugate of that at $(u, v)$, 
\begin{eqnarray}
\overline{V}(u,v) = V(-u,-v).
\end{eqnarray}
The 180\degr-rotation of the image with respect to the origin $(x,y)=(0,0)$ on sky corresponds to flipping the signs of both $(x,y)$ coordinates, i.e., $(x,y)\rightarrow (-x,-y)$.  In the visibility domain, this is mathematically equivalent to flipping the signs of $(u,v)$.  Therefore, the visibility of 180\degr-rotated image is  
\begin{eqnarray}
V(u,v) \rightarrow V(-u,-v) = \overline{V}(u,v).
\end{eqnarray}
Thus, subtracting the 180\degr-rotated image is mathematically equivalent to setting the real part to zero and doubling the value of the imaginary part as follows
\begin{eqnarray}
V(u,v) - V(-u,-v) = V(u,v) - \overline{V}(u,v) = 2{\rm Im}(u,v).
\end{eqnarray}
}
In other words, the real part contains information on both the symmetric and asymmetric structures of objects, whereas the imaginary part contains only information on the asymmetries. Therefore, by synthesizing the image with only the imaginary part, we selectively remove symmetric structures and reveal only asymmetric structures. We determined the minimum r.m.s.\ in the central region of the images by shifting the images relative to the new GAIA phase center in the visibility domain by a phase shift of $\exp\left[2\pi i \left(u\,\Delta {\rm R.A.} +v\,\Delta {\rm DEC}\right)\right]$, where $u$ and $v$ are the spatial frequencies and $\Delta$R.A. and $\Delta$DEC are the shift values. \jh{Figure}~\ref{figA:long} shows dust continuum images synthesized with only the imaginary part, including the image with the minimum r.m.s. We found shift values for the minimum r.m.s.\ ($\Delta$RA, $\Delta$DEC) of ($-$28~mas, $-$23~mas) relative to the phase center determined by GAIA. The final phase center in the ICRS coordinate is (4h30m51.7325, $+$24d41m47.152).

\section{$^{12}$CO~$J=3-2$ channel maps}\label{secA:channel}

\begin{figure}[ht!]
\begin{centering}
\includegraphics[clip,width=\linewidth]{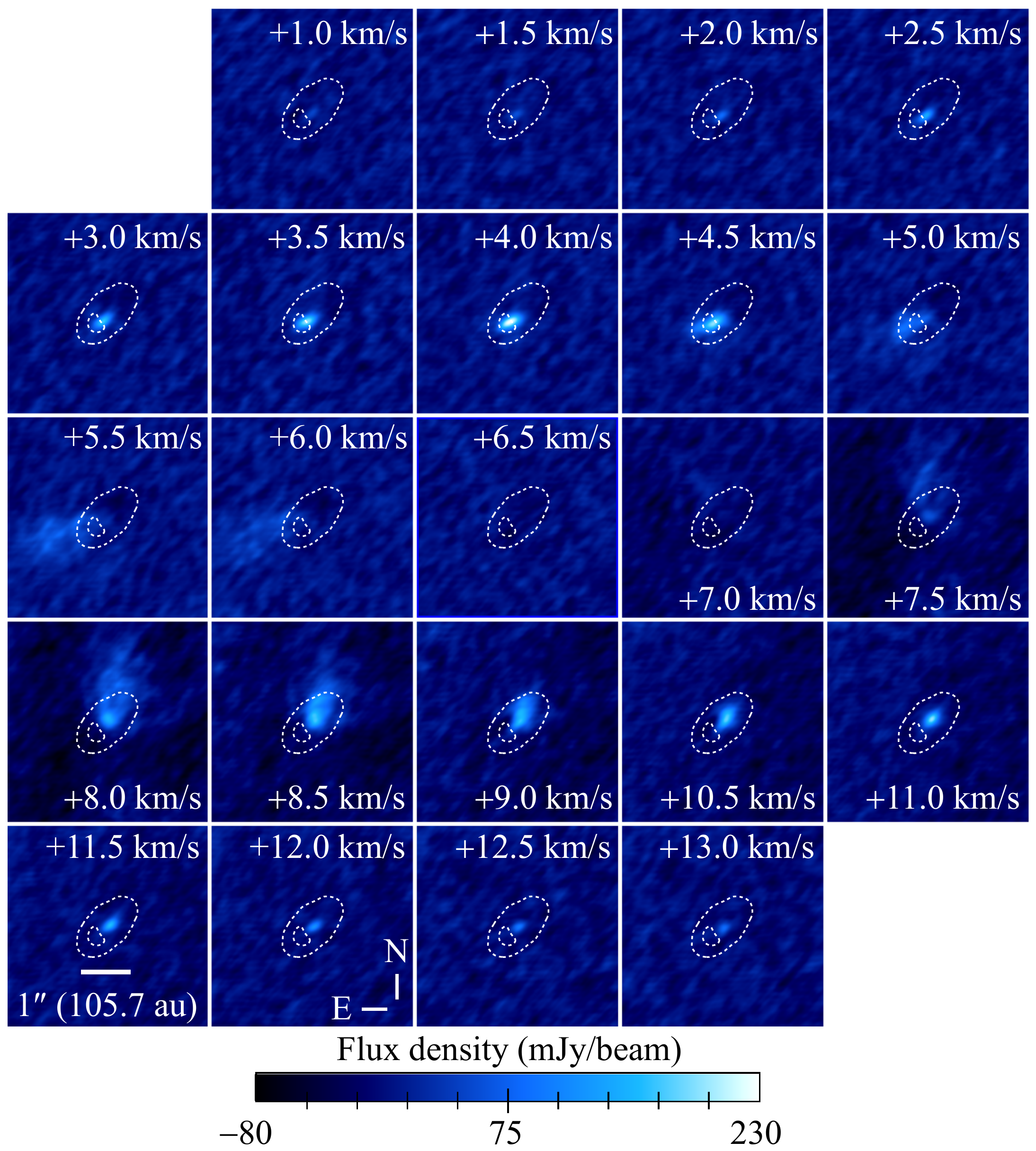}
\end{centering}
\caption{Channel maps of $^{12}$CO~$J=3-2$. The r.m.s noise at the 0.5~km/s bin is 9.56~mJy/beam with a beam size of 283~$\times$~156~mas at a PA of $-$50.2\degr.}\label{figA:channel}
\end{figure}

Figure~\ref{figA:channel} shows the $^{12}$CO~$J=3-2$ channel maps at $+$1.0 to $+$13.0~km/s.

\section{Trace plots in MCMC calculations}\label{secA:trace2}

\begin{figure}[ht!]
\centering
\includegraphics[clip,width=170mm]{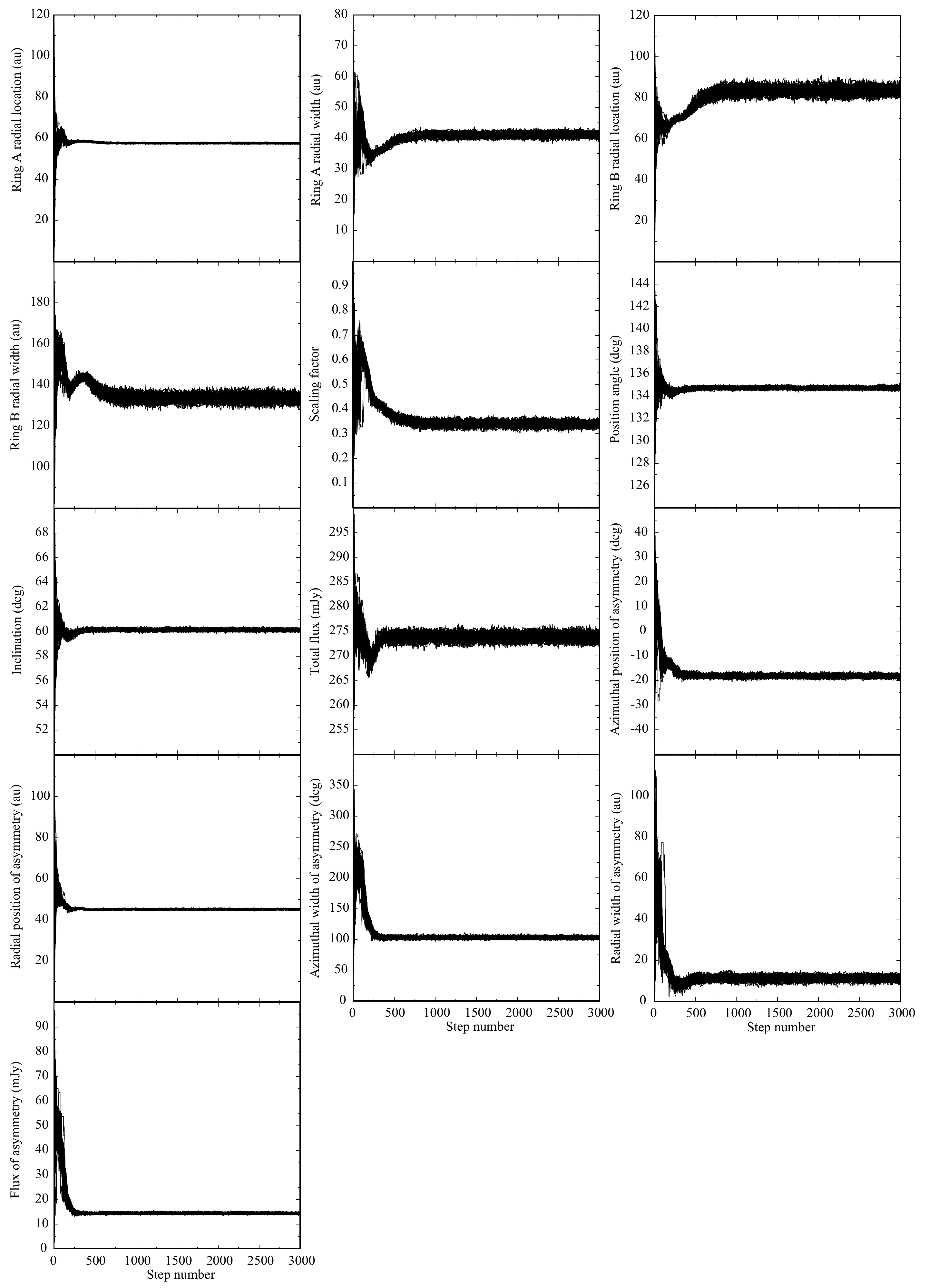}
\caption{Trace plots for 100~walkers for 13 parameters in the model. The initial 1,000~steps are set as the burn-in phase and are discarded in the histograms of the marginal distributions of the MCMC posteriors in Figure~\ref{figA:corner2}.
}\label{figA:trace2}
\end{figure}

Figure~\ref{figA:trace2} shows trace plots for 100~walkers for 13 parameters in the model in our visibility analyses (\S~\ref{sec:mcmc}). The burn-in phase is set as the initial 1,000~steps.

\section{Histogram of the marginal distributions for the MCMC posteriors}\label{secA:corner2}

\begin{figure}[ht!]
\begin{centering}
\includegraphics[clip,width=\linewidth]{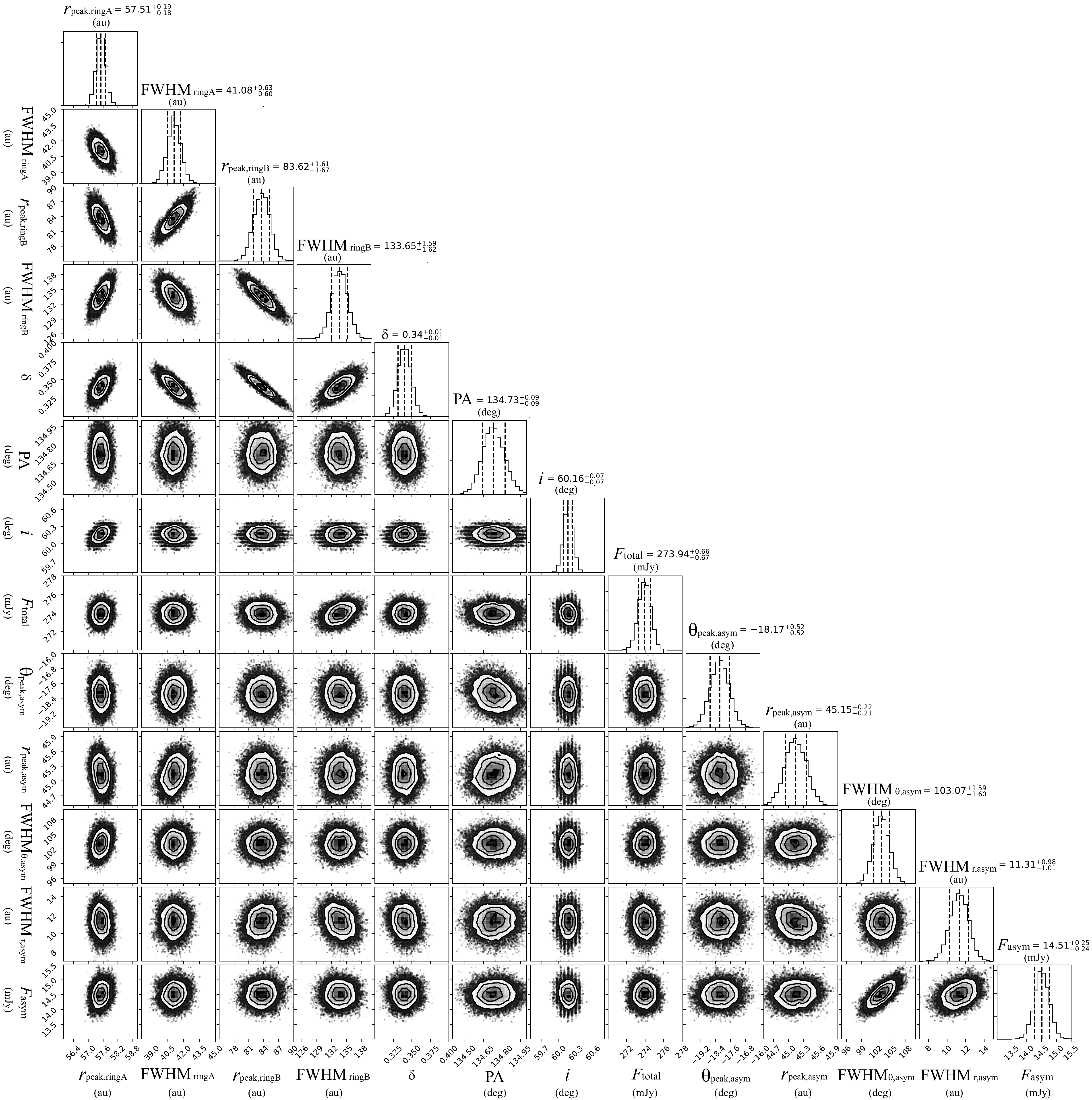}
\end{centering}
\caption{Corner plot of the MCMC posteriors calculated in visibility analyses in \S~\ref{sec:mcmc}. The histograms on the diagonal are marginal distributions of 13 free parameters. The parameter ranges in each parameter are described in Table~\ref{tab:mcmc}. The vertical dashed lines in the histograms represent the median values and the 1~$\sigma$ confidence intervals for parameters computed from the 16th and 84th percentiles. The off-diagonal plots show the correlation for corresponding pairs of parameters.}\label{figA:corner2}
\end{figure}

Figure~\ref{figA:corner2} shows a corner plot of the MCMC posteriors calculated in visibility analyses in \S~\ref{sec:mcmc}.

\jh{
\section{Observed and modeled visibilities}\label{secA:vis}

Figure~\ref{figA:vis} shows observed and modeled visibilities in our visibility analyses (\S~\ref{sec:mcmc}).

\begin{figure}[ht!]
\centering
\includegraphics[clip,width=\linewidth]{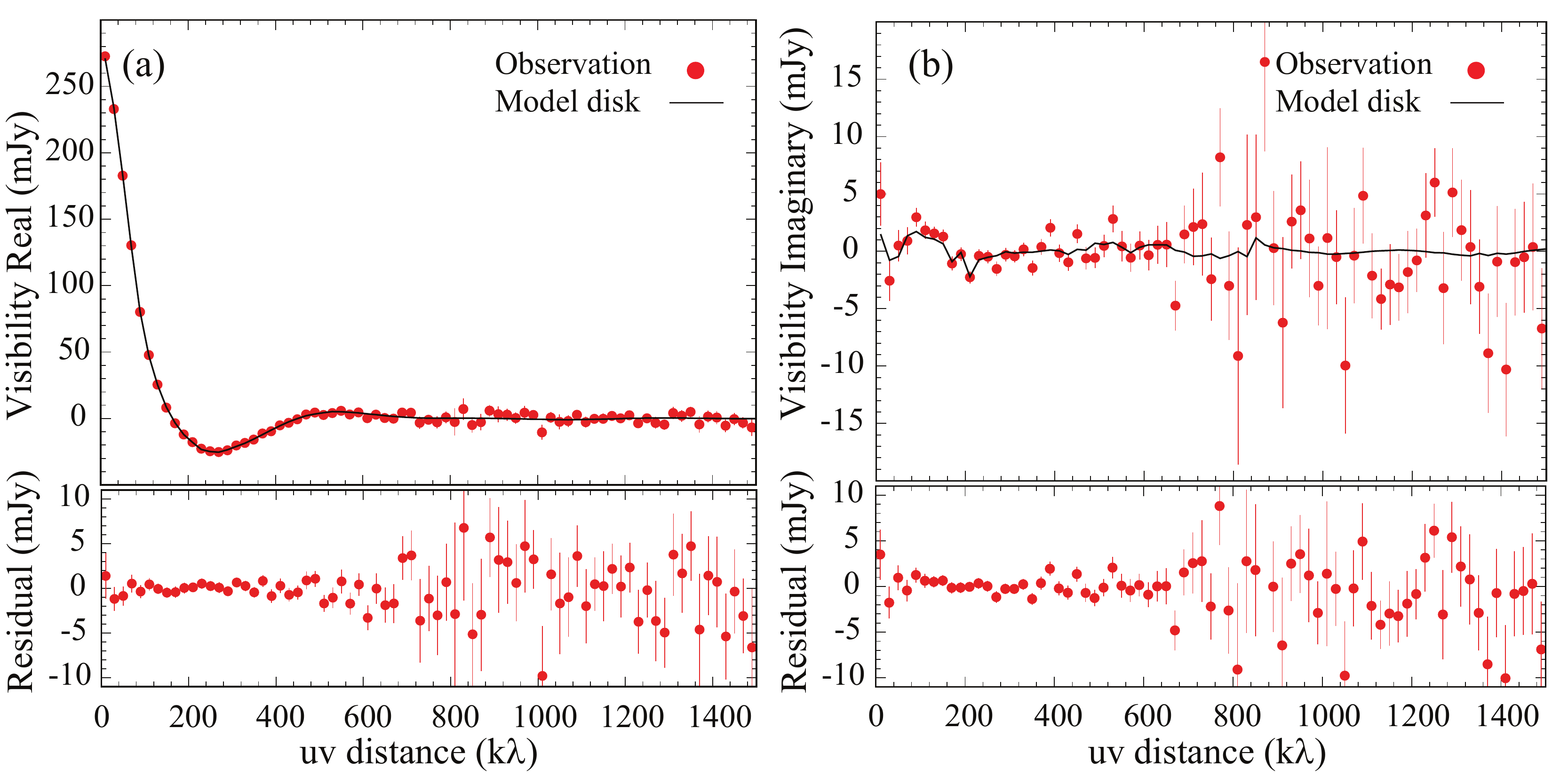}
\caption{Real (a) and imaginary (b) parts of the visibilities for the observations (red dots) and the best-fit model (black line) in the top panel in visibility analyses in \S~\ref{sec:mcmc}. The bottom panel shows residual visibilities between observations and the best-fit model. The reduced-$\chi^{2}$ is 1.91.
}\label{figA:vis}
\end{figure}
}

\section{Weight values and standard deviations in the real and imaginary parts}\label{secA:weight}

\begin{figure}[ht!]
\begin{centering}
\includegraphics[clip,width=\linewidth]{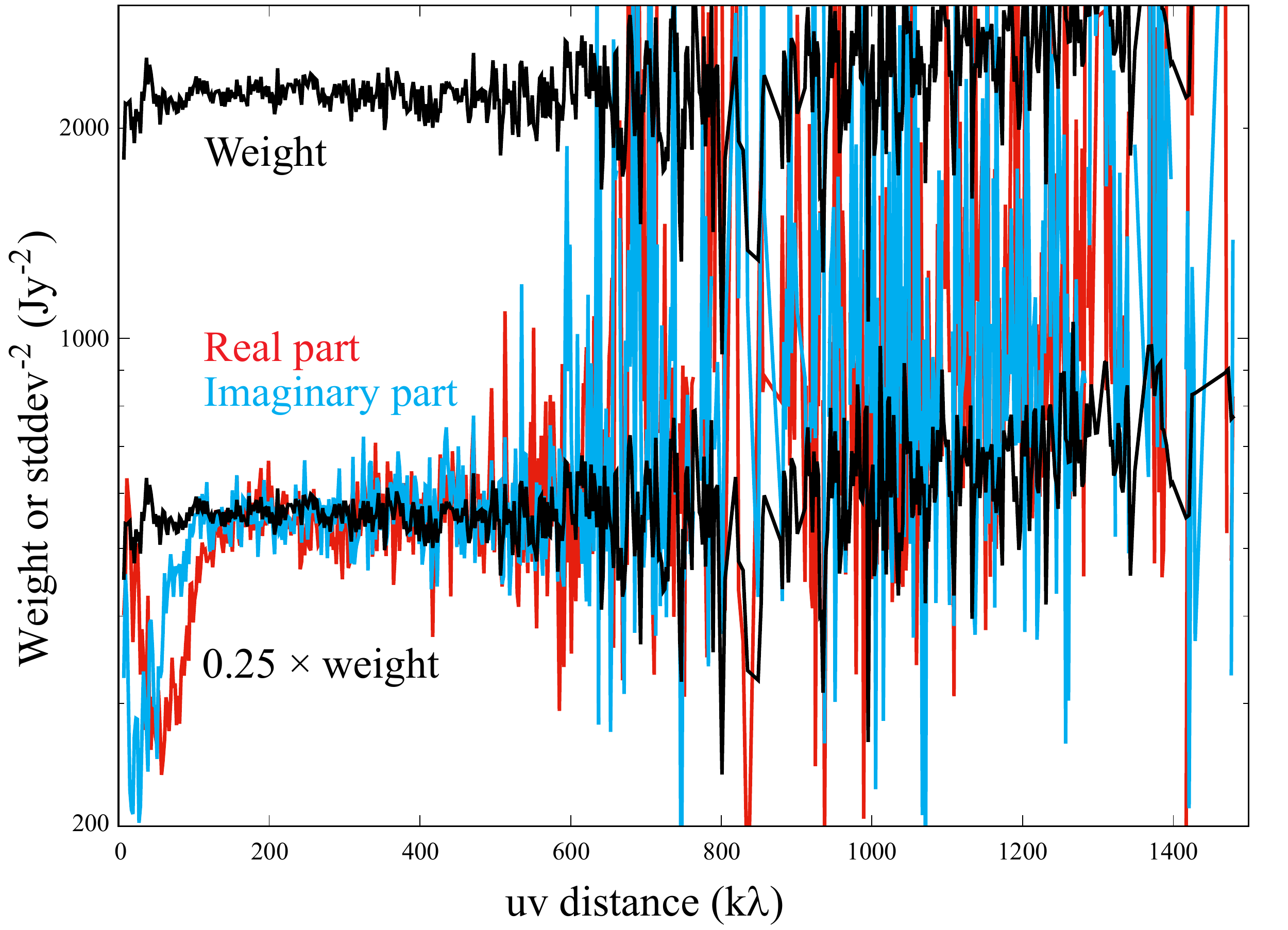}
\end{centering}
\caption{Comparison of weights and standard deviations (stddev) in real and imaginary parts. The weights are typically 4.0$\times$ higher than 1/stddev$^{2}$ of the real and imaginary parts at 100--600~k$\lambda$. 
}\label{figA:weight}
\end{figure}

Figure~\ref{figA:weight} shows a comparison of the values of weights and standard deviations in the real and imaginary parts. \jh{The values of weights ($W_j$) are provided in the measurement set of ALMA data while} the standard deviations (stddev) are calculated in each 2~k$\lambda$~bin in the deprojected visibilities of the real and imaginary parts with $i=$ 60$\degr$ and PA $=$ 135$\degr$. We found that the weights are typically 4.0$\times$ higher than 1/stddev$^{2}$ of the real and imaginary parts at $\sim$100--600~k$\lambda$. 

\bibliography{ref}{}
\bibliographystyle{aasjournal}

\end{document}